\documentclass[prb,twocolumn,showpacs,floats,eqsecnum,amsmath,amssymb]{revtex4}
\usepackage[dvips]{graphicx}
\begin{document}

\title{Tunable Surface Plasmon Polaritons in a Weyl Semimetal Waveguide}

\author{S. Oskoui Abdol, A. Soltani Vala, B. Abdollahipour\footnote{Corresponding
author,
\\Email Address: b-abdollahi@tabrizu.ac.ir}}

\address{Department of condensed matter physics, Faculty of
Physics, University of Tabriz, Tabriz 51666-16471, Iran}
\date{\today}
\begin{abstract}
Weyl semimetals have recently attracted extensive attention due to
their anomalous band structure manifested by topological properties
that lead to some unusual and unique physical properties. We
investigate novel features of surface plasmon polaritons in a slot
waveguide comprised from two semi-infinite Weyl semimetals. We
consider symmetric Voigt-Voigt and Faraday-Faraday configurations
for plasmon polaritons in two interfaces of waveguide and show that
the resulting dispersion is symmetric and the propagation of surface
plasmon polaritons is bidirectional. We introduce exotic and novel
asymmetric structures making use of difference in magnitude or
orientation of chiral anomalies in two Weyl semimetals in both Voigt
and Faraday configurations. These structures show a tremendous
nonreciprocal dispersion and unidirectional propagation of surface
plasmon polaritons. Moreover, we study an hybrid configuration of
Voigt-Faraday for surface plasmon polartions in two interfaces of
the waveguide. We find that this structure possesses unique futures.
It shows surface plasmon polariton modes with unidirectional
propagation above the bulk plasmon frequency. Furthermore, we find a
surface plasmon polariton band which admits the Voigt and Faraday
features simultaneously. Also, we show that the waveguide thickness
and the chemical potential of the Weyl semimetals can be used as a
fine-tuning parameters in these structures. Our findings may be
employed in optical devices which exploit the unidirectional surface
plasmon propagation features.
\end{abstract}
\maketitle
\section{Introduction}
The recent discovery of the topological insulators
(TIs)\cite{Kane05,Hasan10} has led to a surge of interest in
topological properties of the electronic band structure of
crystalline materials. TIs exhibit a bulk gap, but gapless surface
states protected by topology. Weyl semimetals (WSMs) being a
non-trivial phase of matter have recently attracted extensive
attention\cite{Armitage18}. This interest is due to their anomalous
band structure, which is manifested by topological
properties\cite{Murakami07} and protected Fermi arc surface
states\cite{Wan11}. WSMs possess band structure touching at Weyl
nodes, which appear in pairs and are characterized by linear
dispersion close to the Fermi level\cite{Fang03}. In bulk Dirac
semimetals (BDSs) Weyl nodes are doubly degenerate and are called
Dirac nodes, which requires both time reversal and inversion
symmetries to be conserved. In WSMs, breaking the time reversal
symmetry or inversion symmetry leads to a separation of Weyl nodes
with opposite chirality in momentum or energy, respectively. Weyl
semimetal phase has recently been observed in TaAs \cite{Xu15-1},
NbAs \cite{Xu15-2}, NbP \cite{Shekhar15}, $Eu_2Ir_2O_7$
\cite{Sushkov15} and $WTe_2$\cite{Wu16}. Non-trivial band topology
of WSMs emerges in a number of novel exotic effects such as chiral
anomaly\cite{Parameswaran14}, anomalous Hall
effect\cite{Xu11,Burkov14} and negative
magnetoresistance\cite{Huang15}. Furthermore, chiral anomaly in WSMs
is expected to result in the unusual optical responses due to the
coupling of the electric and magnetic
properties\cite{Zyuzin12,Chen13,Vazifeh13,Ashby13,Ashby14}.

Surface plasmon polaritons (SPPs) are collective electromagnetic and
electronic charge excitations which are confined to the interface of
a metal or semiconductor with a dielectric\cite{Maier07}. SPPs
propagate with wavelengths smaller than the light wavelength in
vacuum and can be employed as a platform for developing novel
plasmonic based optoelectronic devices such as surface plasmon
resonance sensor\cite{Homola99} and scanning near field optical
microscopy\cite{Novotny06}. Application of an external magnetic
field parallel to the interface of a metal or semiconductor with
dielectric leads to a nonreciprocal SPP modes with a unidirectional
propagating electromagnetic
waves\cite{Wallis74,Kushwaha87-1,Kushwaha87-2,Kushwaha87-3}. Also,
unidirectional SPP mode has been reported in the system of two
circularly polarized quantum emitters held above a metal
surface\cite{Downing19}. The unidirectional electromagnetic wave
propagation is the subject of chiral quantum optics which deals whit
propagation-direction-dependent light-matter
interactions\cite{Lodahl19}. Optical devices with the nonreciprocal
SPPs are employed for developing unidirectional optical
circuits\cite{Dotsch05} and the directed excitations in a ring
laser\cite{kravtsov99}.

Recently, SPP modes on the surface of a TI have been investigated
theoretically\cite{Raghu10,Efimkin12,Karch11,Schutky13,Qi14} and
experimentally\cite{Pietro13,Lu19}. SPP modes on TI surface exhibit
the same dispersion relation as those of graphene due to their
identical linear Dirac electronic spectra. On the other hand, the
charge and spin density waves are coupled due to the spin-momentum
locking in the TI, giving rise to spin-coupled surface plasmons or
spin plasmons\cite{Raghu10,Efimkin12}. A ferromagnetic coupling or
an external magnetic field brakes the time reversal symmetry of
surface states in TI and causes a magneto-optical Kerr effect. This
effect gives rise to generation of a novel transverse SPP in
addition to the usual longitudinal one on the surface of a
TI\cite{Karch11,Schutky13,Qi14}. Several studies have been devoted
to investigation of the surface plasmon polaritons in BDSs and WSMs.
The properties of plasmon excitation in BDSs have been studied and
it has been shown that these excitations are
universal\cite{Kharzeev15}. It has been shown that in frequencies
lower than the Fermi energy the metallic response is dominated in a
BDS film and manifests in the existence of the SPPs, but at higher
frequencies the dielectric response is dominated and it behaves as a
dielectric waveguide\cite{Kotov16}. SPPs behavior in the interface
of a Bulk WSM and a dielectric has been studied for different
orientations of the Weyl nodes separation vector and the SPP
propagation direction\cite{Hofmann16}. It has been shown that the
SPP dispersion depends on the Weyl nodes separation in energy or
momentum space and for a time reversal broken WSM the Weyl nodes
separation acts as an effective external magnetic field. Moreover,
in the Voigt configuration SPP has a nonreciprocal unidirectional
dispersion. In the Faraday and perpendicular configurations the SPP
dispersion develops a gap at an intermediate frequency region.
Further, studies have been performed for nonreciprocal propagation
of SPP in Weyl semimetal thin films. The existence of giant
nonreciprocal waveguide electromagnetic modes in WSM thin films in
the Voigt configuration have been predicted\cite{Kotov18}. Also, it
has been shown that the SPP dispersion and its nonreciprocal
property can be controlled by fine-tuning of the thickness of WSM
thin film and dielectric contrast of the outer
insulators\cite{Tamaya18}. Recently, the generation of SPPs at
visible wavelengths in the WSM $WTe_2$ has been
reported\cite{Tan18}. The nonreciprocal unidirectional propagation
of the electromagnetic modes has been studied extensively in the
context of magnetoplasmons in dielectric waveguides with ferrite
substrate and films of magnetic
dielectrics\cite{Chiu72,Hartstein74,Kushwaha01}.

The inherent properties of the SPPs on the surface of WSM are caused
by its intrinsic topological properties without need to application
of high external magnetic fields (up to several tens of tesla).
These topological properties fixes strength of the coupling of the
electric and magnetic properties of WSMs through the chiral magnetic
effect which depends on the separation of the Weyl nodes in momentum
space. The transverse or Hall conductivity in these materials which
is responsible for inhomogeneous optical responses of WSMs is
estimated to be several orders of magnitude larger than typical
magnetic dielectrics\cite{Kotov18,Tamaya18}. Therefore, the
intrinsic topological properties of WSMs provide the opportunity to
stable and efficient control of SPP propagation at the interface of
these materials. Motivated by these intriguing properties of SPPs at
the interface of a WSM, we intend to study SPP dispersion and
localization in a WSM slot waveguide. Strong and intrinsic chiral
magnetic effect in WSMs provides the opportunity to consider more
achievable configurations in a WSM slot waveguide. We study
symmetric Voigt-Voigt and Faraday-Faraday waveguides and show that
as it is expected the SPP dispersion in these structures are
reciprocal. Also, we find that a robust nonreciprocallity and
unidirectional propagation of SPPs in Voigt-Voigt configuration can
be achieved by contrasting the magnitude or direction of the chiral
magnetic vectors in two WSMs. Further, we analyze the SPPs
dispersion in the hybrid Voigt-Faraday configuration and again
retrieve a giant nonreciprocal SPP propagation for frequencies above
the bulk plasmon frequency. Moreover, we observe some novel and
exotic features such as a SPP dispersion band which inherit
simultaneously Voigt and Faraday configurations properties. Also, we
show that the thickness of the slot waveguide and the chemical
potential of the WSMs can be used as fine-tuning to control the SPPs
propagation in these structures. These fascinating features being
originated from intrinsic topological properties of the WSMs make
them experimentally feasible and on the other hand may be very
important from the practical perspective.

The remainder of the paper is organized as follows: In Sec. \ref{S1}
we give some basic background about the optical responses of WSMs.
Sec. \ref{S2} and its subsections have been devoted to present the
derivation of dispersion relation for Voigt-Voigt, Faraday-Faraday
and Voigt-Faraday configuration and discussing properties of SPP
dispersion in these structures. Finally, we end by giving conclusion
In Sec. \ref{S3}.

\section{The theoretical background}\label{S1}

In the bulk of the WSM the valance and conduction bands touch each
other at the Weyl nodes which appear in pairs with opposite
chiralities. A WSM with broken time reversal symmetry contains two
Weyl nodes with opposite chiralities separated in momentum space,
while for a WSM with broken inversion symmetry Weyl nodes are
separated in energy space. The low energy Hamiltonian in the
vicinity of these points is given by\cite{Vazifeh13},
\begin{equation}
\hat{H} = \chi v_{F}\mathbf{\sigma}\cdot(\mathbf{k} - \mathbf{b}) +
\chi b_{0} ,\label{eq2.1}
\end{equation}
where $v_{F}$ is the Fermi velocity, $\chi=\pm 1$ denotes the
chirality, $\mathbf{k}=(k_x, k_y, k_z)$ is the momentum operator and
$\sigma=(\sigma_x, \sigma_y, \sigma_z)$ is the vector of the Pauli
matrices. $\mathbf{b}$ and $b_{0}$ indicate the separation of two
Weyl nodes in momentum and energy, respectively. The topological
properties of the WSM is explained by $\theta (\mathbf{r},t) =
2(\mathbf{b}.\mathbf{r} - {b_0}t)$ which is referred as
\textit{axion~angle}\cite{Vazifeh13}. For $\mathbf{b}=b_{0}=0$, the
bands are degenerate and the material does not possess topological
properties. It is the case of so called BDS. The axion angle effect
is described by an additional term $L_{\theta}$ in the Lagrangian of
the system\cite{Wilczek87},
\begin{equation}
\mathcal{L}_{em} = \frac{1}{{8\pi }}({\mathbf{E}^2} -
{\mathbf{B}^2}) - \rho\varphi + \mathbf{J}\cdot\mathbf{A} +
L_{\theta} ,\label{eq2.2}
\end{equation}
\begin{equation}
L_{\theta} = -
\frac{\alpha}{4\pi^2}\theta(r,t)\mathbf{E}\cdot\mathbf{B} ,
\label{eq2.3}
\end{equation}
where $\mathbf{E}$, $\mathbf{B}$, $\varphi$ and $\mathbf{A}$ are the
electric field, magnetic field, electric potential and magnetic
vector potential, respectively. Here, the charge and current
densities are denoted by $\rho$ and $\mathbf{J}$. In the above
equation $\alpha$ is a constant called effective fine structure
constant of the WSM. Thus, the resulting Maxwell's equations are,
\begin{eqnarray}
\nabla\cdot\mathbf{E} &=& 4\pi (\rho  + \frac{\alpha
}{2\pi^2}\mathbf{b}\cdot\mathbf{B}) ,\nonumber\\ -
\frac{1}{c}\frac{\partial \mathbf{E}}{\partial t} + \nabla
\times\mathbf{B} &=& \frac{4\pi }{c}[\mathbf{J} - \frac{\alpha
}{2\pi ^2}(c\mathbf{b} \times
\mathbf{E} - b_0\mathbf{B})] ,\nonumber\\
\nabla\times\mathbf{E} &=& - \frac{1}{c}\frac{{\partial
\mathbf{B}}}{{\partial t}}, \nonumber\\ \nabla\cdot\mathbf{B} &=& 0
. \label{eq2.4}
\end{eqnarray}
As a result, the charge and current densities are modified in the
Maxwell's equations by additional terms proportional to
$-\nabla\theta\cdot\mathbf{B}$ and $\nabla\theta\times\mathbf{E} +
\dot\theta\mathbf{B}$, respectively. So the displacement electric
field is given by\cite{Hofmann16},
\begin{equation}
\mathbf{D} = (\varepsilon _{\infty}  + \frac{4\pi i}{\omega }\sigma
)\mathbf{E} + \frac{{i{e^2}}}{{\pi \hbar \omega }}(\nabla \theta )
\times \mathbf{E} + \frac{{i{e^2}}}{{\pi \hbar c\omega }}\dot \theta
\mathbf{B}\ , \label{eq2.5}
\end{equation}
where $\varepsilon_{\infty}$ is the static dielectric constant of
WSM and $\sigma$ is the conductivity. First term of the above
equation represents the displacement field for a normal metals,
while the two last terms originate from chiral anomaly representing
the anomalous Hall effect (AHE) and chiral magnetic effect (CME),
respectively\cite{Armitage18}.

For a WSM with broken time reversal symmetry, the chiral anomaly
causes an anisotropic optical response with the diagonal and off
diagonal terms given by ${\varepsilon}(\omega ) ={\varepsilon
_\infty}(1 - \frac{{\Omega _p}^2}{{\omega ^2}})$ and
${\varepsilon_b}(\omega ) = {\varepsilon _\infty}\frac{\omega
_b}{\omega}$, respectively. Where ${\Omega_p}^2 =
\frac{4\alpha}{3\pi}{(\frac{\mu}{\hbar})^2}$ refers to bulk plasmon
frequency with $\alpha =\frac{e^2}{\hbar {v_f}{\varepsilon _\infty
}}$, $\mu$ chemical potential and ${\omega _b} = 2{e^2}\left| b
\right|/\pi \hbar {\varepsilon _\infty}$. Measurements have been
revealed an ultrahigh mobilities (much higher than the best
graphene) and very small carrier scattering rates for BDSs and
WSMs\cite{Shekhar15,Sushkov15}. This mainly is due to the
crystalline symmetries of these materials and their linear
electronic dispersion relation around the Weyl nodes. Therefore, we
have ignored the effect of the carrier scattering in the dielectric
tensor of WSM and we disregard the effect of loss on SPP propagation
in the subsequent calculations.

To study the SPP localized at the interface of a WSM and a
dielectric, located at  $x-y$ plane, we assume the electric field in
the following form,
\begin{equation}
\mathbf{E} = (E_x,E_y,E_z){e^{i{q_x}x + i{q_y}y}} {e^{ - i\omega
t}}{e^{ - \kappa \left| z \right|}} .\label{eq2.6}
\end{equation}
This electric field decays exponentially away from interface in the
$z$ direction and propagates in the interface along the direction of
$\mathbf{q}=(q_x,q_y)$. The decay constant $\kappa>0$ is obtained by
solving the wave equation,
\begin{equation}
\nabla  \times (\nabla  \times\mathbf{E}) =  -
\frac{1}{{{c^2}}}\frac{{{\partial ^2}}}{{\partial {t^2}}}\mathbf{D}
.\label{eq2.7}
\end{equation}
Substituting the electric field given by Eq.(\ref{eq2.6}) in the
wave equation (\ref{eq2.7}) leads to a system of three linear
equations,
\begin{eqnarray}
&\hat{ M}\cdot\mathbf{E}= 0\ ,\\\nonumber &\hat{M} = \left(
{\begin{array}{*{20}{c}}
{q_y^2 - \kappa _j^2}&{ - {q_x}{q_y}}&{ \mp i{q_x}{\kappa _j}}\\
{ - {q_x}{q_y}}&{q_x^2 - \kappa _j^2}&{ \mp i{q_y}{\kappa _j}}\\
{ \mp i{q_x}{\kappa _j}}&{ \mp i{q_y}{\kappa _j}}&{{q_x}^2 +
{q_y}^2}    \end{array}} \right) - \frac{{{\omega ^2}}}{{{c^2}}}\hat
\varepsilon (\omega ) ,\label{eq2.8}
\end{eqnarray}
with the positive sign for the dielectric side and the negative one
for the WSM side, $\hat{\varepsilon}$ denotes the dielectric tensor
and $j=1,2$ refers to the different regions. The zeroes of the
determinant of coefficient matrix ($\hat{M}$) yield the decay
constants. There are different solutions for decay constant on the
WSM side depending on the relative direction of the vectors
$\mathbf{q}$ and $\mathbf{b}$\cite{Hofmann16}. The dispersion
relation is obtained by applying boundary conditions at the
interface which are continuity of the tangential components of the
electric and magnetic fields.

SPPs are categorized in three different configurations according to
the relative orientation of the vectors $\mathbf{b}$ and
$\mathbf{q}$ whit respect to the surface: (a) Voigt geometry:
$\mathbf{b}$ parallel to the surface, but perpendicular to
$\mathbf{q}$ (b) Faraday geometry: $\mathbf{b}$ parallel to the
surface and $\mathbf{q}$ (c) Perpendicular geometry: $\mathbf{b}$
perpendicular to the surface and $\mathbf{q}$. Here we consider the
Voigt and Faraday configurations and their composition.
%
\section{Surface Plasmon Polaritons in a Slot Waveguide}\label{S2}
\begin{figure}
\centerline{\includegraphics[width=8cm]{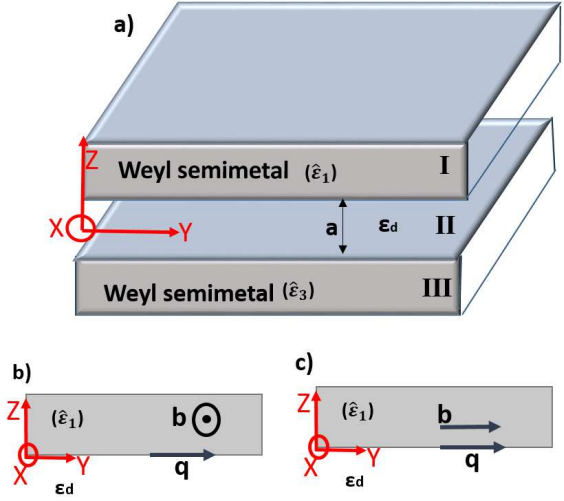}} \caption{(a)
Schematic of a slot waveguide constructed of two semi-infinite WSMs
connected by a dielectric layer with thickness $a$ and dielectric
constant $\varepsilon_{d}$. WSMs are considered in two different
Voigt and Faraday configurations as illustrated in figures (b) the
Voigt configuration and (c) the Faraday configuration.}
\label{system-model}
\end{figure}

%
The schematic structure of the system studied in the present paper
has been depicted in Fig.\ref{system-model}. This structure is
composed of two semi-infinite layers of the WSM and an insulator
layer with  dielectric constant $\varepsilon_{d}$ and thickness $a$
in the middle. Optical properties of WSM layers are determined by a
dielectric tensor which is intrinsic property of them in contrast to
the semiconductor magneto optic materials in which the anisotropic
response originates from an external magnetic field. The system of
coordinates has been chosen so that the interfaces of WSMs and the
middle dielectric to lie in the x-y plane and the $z$ axis to be
perpendicular to the interfaces. The relative orientation of the
vectors in Voigt and Faraday Configurations has been shown in
Fig.\ref{system-model} (b) and (c) as well. In the following
sections we consider different situations combining Voigt and
Faraday configurations and demonstrate the properties of the SPPs
supported by these structures.
%
%
%
\subsection{Surface plasmon polaritons in Voigt-Voigt waveguide}
In this section we consider WSMs to be in the Voigt configuration.
We assume the SPP propagation is in the $y$ direction
$\mathbf{q}=(0,q_y,0)$ and for both WSMs we take $\mathbf{b}$ to be
parallel with $x$ axis, $\mathbf{b}=(b,0,0)$. Thus the dielectric
tensors of mediums I and III (see Fig. \ref{system-model}) are given
by,
\begin{equation}
\begin{array}{*{20}{c}}
{\hat{\varepsilon}_{V,i}(\omega ) = \left(
{\begin{array}{*{20}{c}} {\varepsilon_i}&0&0\\
0&{\varepsilon_i}&{i\varepsilon _{bi}}\\
0&{ -i \varepsilon_{bi}}&{\varepsilon_i}
\end{array}} \right)}&{,i \equiv 1,3}
\end{array}\label{eq3.1}
\end{equation}
Substituting the dielectric tensor in the wave equation, Eq.
(\ref{eq2.7}), we obtain the coefficient matrix $\hat{M}$ as,
\begin{equation}\small{
{\hat{M}_{V,i}} = \left( {\begin{array}{*{20}{c}}
{{q^2} - {k_i}^2 - {k_{0} ^2}\varepsilon_i}&0&0\\
0&{ - {k_i}^2 - {k_{0} ^2}\varepsilon_i}&{ \mp
iq{k_i} - i{k_{0}^2}\varepsilon _{bi}}\\
0&{ \mp iq{k_i} + i{k_{0} ^2}\varepsilon _{bi}}&{{q^2} - {k_{0}
^2}\varepsilon_i}
\end{array}} \right) }.\label{eq3.2}
\end{equation}
Setting the determinant of $\hat{M}$ to zero, we obtain two
solutions $k_{i}^{+}$ and $k_{i}^{-}$ for decaying wave vector:
\begin{equation}
{\begin{array}{*{20}{c}}k_i^+ = \sqrt{q^2 - {k_{0} ^2}\varepsilon_i}
\\k_i^- =\sqrt{ q^2 - {k_{0}^2}\varepsilon_{vi}}\ ,
\end{array}}\label{eq3.3}
\end{equation}
which are attributed to TE and TM modes, respectively. Where
$\varepsilon_{vi} = (\varepsilon_i^2 -
\varepsilon_{bi}^2)/\varepsilon_i$ is the Voigt dielectric constant
and $k_{0}=\omega/c$ is the vacuum wave vector. Since TE polarized
waves are not affected by the chiral anomaly, similar to the magneto
plasmons in the semiconductor magneto optic material\cite{Wallis74},
thus we consider only TM mode. A same procedure also gives the wave
equation in the isotropic medium \textbf{II}, if  we set diagonal
elements of the dielectric tensor as
$\varepsilon_{xx}=\varepsilon_{yy}= \varepsilon_{zz}=\varepsilon _d$
and off diagonal elements to zero. Therefore, the decay constant in
the dielectric layer is obtained as $k_2=\sqrt{q^2 -
{k_{0}}^2{\varepsilon_d}}$. The electric field takes the following
form throughout the structure,
\begin{equation}
\mathbf{E}(r,t) = \mathbf{E}(z){e^{i(qy - \omega t)}}\label{eq3.4}
\end{equation}
The field amplitude in regions \textbf{I} ($z\geq a/2$), \textbf{II}
($-a/2< z< a/2$) and \textbf{III} ($z\leq-a/2$) are expressed as
follows:
\begin{equation}\small{
\left\{ \begin{array}{l} {\mathbf{E}_1} =
\left({\begin{array}{*{20}{c}}0&,{{E_{1y}}{e^{ - {k_1^{-}}(z -
a/2)}}}&,{{\beta _1}{E_{1y}}{e^{ - {k_1^{-}}(z - a/2)}}}
\end{array}} \right)\\{\mathbf{E}_2} = \left({\begin{array}{*{20}{c}}
0&,{{E_{2y}^+ }{e^{ + {k_2}z}} + {E_{2y}^-}{e^{ -
{k_2}z}}}&,{\beta_2 ({E_{2y}^+}{e^{ + {k_2}z}} - {E_{2y}^-}{e^{ -
{k_2}z}})}\end{array}} \right)\\{\mathbf{E}_3} = \left(
{\begin{array}{*{20}{c}} 0&,{{E_{3y}}{e^{ + {k_3^{-}}(z -
a/2)}}}&,{{\beta _3}{E_{3y}}{e^{ + {k_3^{- }}(z -
a/2)}}}\end{array}} \right)
\end{array} \right.} \label{eq3.5}
\end{equation}
where ${\beta_1} = \frac{{{E_{1z}}}}{{{E_{1y}}}} =- i \frac{{( -
q{k_1^{ - }} + {k_{0} ^2}\varepsilon_{b1})}}{{{q^2} - {k_{0}
^2}\varepsilon_1}}$, $\beta_2  = \frac{{{E_{2z}}}}{{{E_{2y}}}} =  -
\frac{{ iq{k_2}}}{{{q^2} - {k_{0} ^2}{\varepsilon _d}}}$ and ${\beta
_3} = \frac{{{E_{3z}}}}{{{E_{3y}}}} = - i \frac{{(  q{k_3^{ - }} +
{k_{0}^2}\varepsilon_{b3})}}{{{q^2} - {k_{0} ^2}\varepsilon_3}}$.
Imposing the continuity of tangential field components
($E_{y},H_{x}$) as boundary conditions in two interfaces ($z=\pm
a/2$) yields the following equations,
\begin{equation}
\begin{array}{l}{e^{{k_2}a/2}}{E_{2y}^+ } +
{e^{ - {k_2}a/2}}{E_{2y}^-} = {E_{1y}} ,\\
{e^{{k_2}a/2}}{E_{2y}^+ } - {e^{ - {k_2}a/2}}{E_{2y}
^-} = ( - \frac{{{k_2\varepsilon _{w1}}}}{{{k^{-}_{1}\varepsilon_d}}}){E_{1y}} ,\\
{e^{ - {k_2}a/2}}{E_{2y}^+ } + {e^{ + {k_2}a/2}}{E_{2y}^-} = {E_{3y}} ,\\
{e^{ - {k_2}a/2}}{E_{2y}^+ } - {e^{ + {k_2}a/2}}{E_{2y}^- } = ( -
\frac{{{k_2\varepsilon _{w3}}}}{{{k^{-}_{3}\varepsilon_d}}}){E_{3y}}
. \end{array}\label{eq3.6}
\end{equation}
Here $\varepsilon_{w1} = k^{-}_{1}\frac{-q\varepsilon_{b1} +
k^{-}_{1}\varepsilon_1}{q^2 - k_{0}^2\varepsilon_1}$ and
$\varepsilon_{w3} = k^{-}_{3}\frac{-q\varepsilon_{b3} -
k^{-}_{3}\varepsilon_3}{q^2 - k_{0}^2\varepsilon_3}$. To obtain
nonzero solutions for components of the electric fields ${E_{1y}}$,
${E_{2y}^{+}}$, ${E_{2y}^{-}}$ and ${E_{3y}}$, determinant of the
coefficient matrix of above equations should be zero. This leads to
the dispersion relation of the Voigt-Voigt waveguide,
\begin{equation}\small{
\frac{\varepsilon_{d}}{k_{2}}\left(-\frac{k_{3}^{-}}{\varepsilon_{w3}}
+ \frac{k_{1}^{-}}{\varepsilon_{w1}}\right)+\left[1 - \left(
\frac{\varepsilon_{d}}{k_{2}}\right)^2
\left(\frac{k_{1}^{-}k_{3}^{-}}{\varepsilon_{w1}\varepsilon_{w3}}
\right) \right]\tanh(k_2 a) = 0} .\label{eq3.7}
\end{equation}
%
%
\begin{figure}
\centerline{\includegraphics[width=10.5cm]{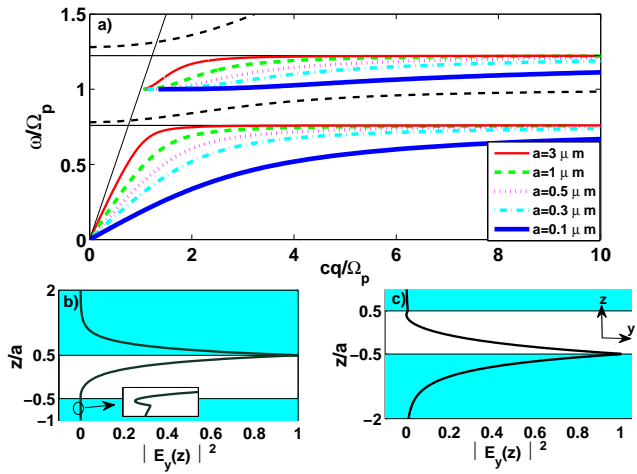}} \caption{a)
Surface plasmon polariton dispersion of the symmetric slot waveguide
in Voigt configuration with $\omega_b=0.5~ \Omega_p$,
${\varepsilon_\infty}=13$, $E_f=0.15~\textit{eV}$, $v_f=10^6~m/s$,
${\Omega_p}=6.0918\times 10^{13} s^{-1}$ and ${\varepsilon_d}=1$. In
this plot, the bulk plasmon dispersion is indicated by black dashed
line. The horizontal black solid thin lines indicate the asymptotic
frequencies. The normalized electric field intensities of the SPP
for thickness of $3 \mu m$ and at frequencies $74.137~THz$ and
$45.768~THz$ for b) the higher band and c) the lower band.}
\label{fig2}
\end{figure}
%
\begin{figure}
\centerline{\includegraphics[width=9cm]{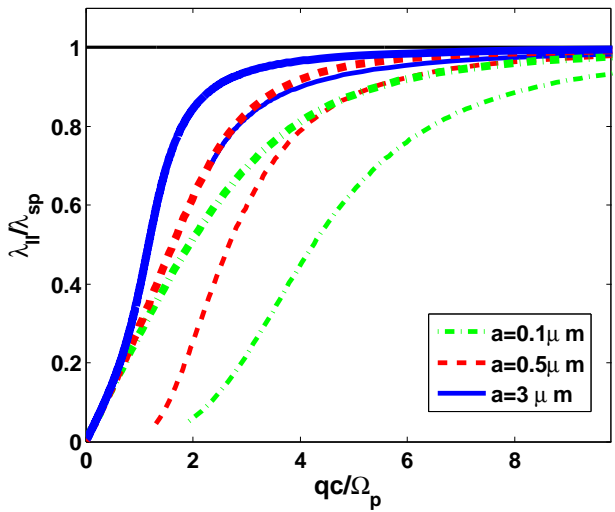}} \caption{The
normalized localization length vs SPP wave vector of the symmetric
slot waveguide in Voigt configuration with $\omega_b=0.5~ \Omega_p$.
Here $\lambda_{ll}$ for  lower and higher SPP bands are indicated by
thick and thin lines, respectively.} \label{fig3}
\end{figure}
%
{\it Symmetric Voigt-Voigt waveguide}: To study SPP in the symmetric
waveguide, we set $\mu_1=\mu_2=\mu$ and $b_1=b_2=b$ which leads to
the following quadratic dispersion equation in q:
\begin{equation}\small{
\frac{{{\varepsilon _d}}}{{{k_2}}}\left(\frac{{2\varepsilon k({q^2}
- \varepsilon {k_0}^2)}}{{{{(\varepsilon k)}^2} - {{({\varepsilon
_b}q)}^2}}}\right) + \left[ {1 + {{(\frac{{{\varepsilon
_d}}}{{{k_2}}})}^2}\frac{{{{({q^2} - \varepsilon
{k_0}^2)}^2}}}{{{{(\varepsilon k)}^2} - {{({\varepsilon _b}q)}^2}}}}
\right]\tanh ({k_2} a) = 0 },\label{eq3.8}
\end{equation}
where we have defined $\varepsilon=\varepsilon_1=\varepsilon_3$ and
$\varepsilon_{b}=\varepsilon_{b1}=\varepsilon_{b3}$ and
$k=k_{1}^{-}=k_{3}^{-}$. Therefore, the nonreciprocal effect
reported in Ref \cite{Hofmann16} for the single interface of a WSM
and a dielectric disappears in the symmetric waveguide structure due
to the symmetry consideration and mixing of the SPPs at two
interfaces.

In this case dispersion curves of the SPP have been plotted for
different thicknesses $a=0.1, 0.3, 0.5, 1, 3~\mu m$ of the
dielectric medium in Fig.\ref{fig2}. In numerical calculation, we
adopt a typical values for the parameters of WSMs as
${\varepsilon_\infty}=13$, ${\omega_b}/{\Omega_p}=0.5$,
$E_f=0.15~eV$, $v_f=10^6~m/s$, ${\Omega_p}=6.0918\times 10^{13}
s^{-1}$, measured for $Eu_2Ir_2O_7$\cite{Sushkov15,Hofmann16} and
${\varepsilon_d}=1$. As it can be seen from the figure, the
dispersion curves of the SPP are composed of two bands which can be
attributed to two distinct ways of the electron
oscillations\cite{Hu12}. One of the bands appears below the bulk
plasmon frequency and the other one above it, which hereafter we
call them the lower and higher bands, respectively. The lower band
starts from the origin and deviates from the light line of the
dielectric layer and then it approaches to the asymptotic frequency
in the large wave vectors. The higher band starts from the light
line at $\omega=\Omega_{p}$ and immediately tends to its asymptotic
frequency. As it has been shown in Fig. \ref{fig2}(a), by decreasing
the thickness of the dielectric layer both SPP bands shift to the
lower frequencies. In order to show the profile of the fields, we
have displayed in Figs. \ref{fig2}(b) and \ref{fig2}(c) the
normalized $y$ component of the electric field intensity for lower
and higher bands of SPP at frequencies $45.768~THz$ and $74.137~THz$
for the thickness $3~\mu m$, respectively. It is obvious that the
lower band of SPP has been highly localized at the lower interface
($z=-a/2$), while the higher band of SPP has been confined mostly to
the upper interface ($z=+a/2$).

In the non-retarded limit ($|q|\gg k_{0}$), where $k=k_{2}=q$, the
dispersion relation is reduced to,
\begin{equation}
2\varepsilon_{d}\varepsilon+
(\varepsilon_{d}^2+\varepsilon^{2}-\varepsilon_{b}^{2} )\tanh(|q|
a)=0 .\label{eq3.9}
\end{equation}
The asymptotic values of the dispersion curves denoted by thin black
lines in Fig. \ref{fig2}(a) are solutions of Eq. (\ref{eq3.9}) for
$|q|\rightarrow\infty$,
\begin{equation}
\omega_{as}^{v}=\frac{\sqrt{\varepsilon_{\infty}^{2}\omega_{b}^{2}+
4\varepsilon_{\infty}
\Omega_{p}^{2}(\varepsilon_{d}+\varepsilon_{\infty})}\pm
\varepsilon_{\infty}\omega_{b}}{2(\varepsilon_{d}+\varepsilon_{\infty})}
, \label{eq3.10}
\end{equation}
where positive (negative) sign is associated for higher (lower)
bands. As it is obvious, for $\omega_{b}=0$ the asymptotic
frequencies of two bands are identical and they merge to one band,
which means the WSM converts to a BDS. The normalized localization
length ($\lambda_{ll}/\lambda_{sp}=q/k$), which characterizes decay
of the electric field component of SPP away from the interface is
plotted as a function of wave vector for different thicknesses of
the dielectric layer in Fig. \ref{fig3}, where thick and thin lines
correspond to the normalized localization length for lower and
higher bands, respectively. As we expect, in this configuration the
numerical results reveal that the decay constants are real quantity
for all frequencies. As this figure shows, the localization length
for both bands decreases by decreasing the thickness of the
waveguide for intermediate wave vectors. For large wave vectors
$qc/\Omega_{p}\gg1$, localization length in all cases approaches to
asymptotic value $\lambda_{sp}=2\pi/q$.

\begin{figure*}
\centerline{\includegraphics[width=18cm]{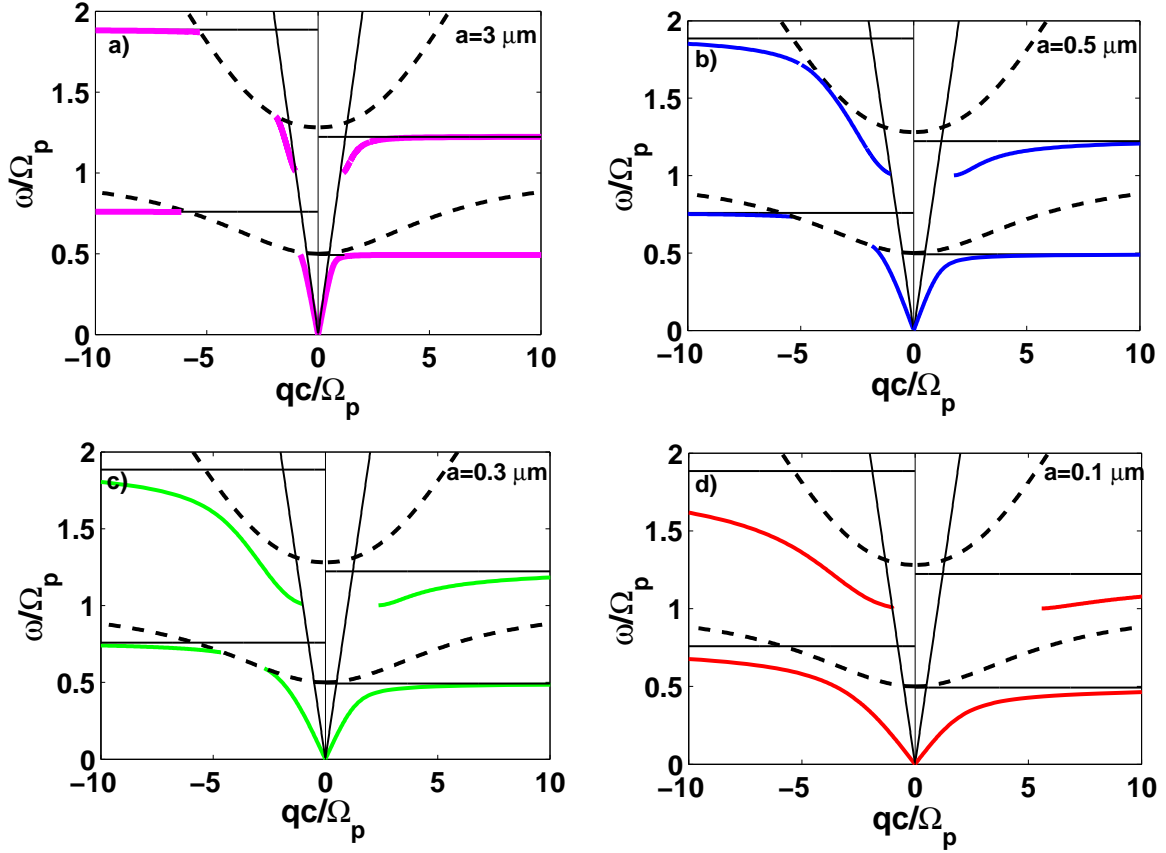}} \caption{SPP
dispersion curve as a function of wave vector for the asymmetric
Voigt-Voigt waveguide with $\omega_{b1}=0.5~ \Omega_p$ and
$\omega_{b3}=1.5~ \Omega_p$ and the other parameter are similar to
Fig.\ref{fig2}. The black dashed lines show the bulk plasmon
dispersion and the horizontal black solid thin lines indicate the
asymptotic frequencies.} \label{fig4}
\end{figure*}

{\it Asymmetric Voigt-Voigt waveguide}: Let us to consider an
asymmetric waveguide constructed by two WSMs in the Voigt
configuration having different $\mathbf{b}$ vectors,
$\mathbf{b}_1\neq\mathbf{b}_3$. We study two distinct cases of the
difference in magnitude and orientation of the $\mathbf{b}$ vectors.
First, we consider the case of contrast in the magnitude of the
chiral anomalies $b_1\neq b_2$. Thus, two WSMs have similar
$\hat{M}$ matrix with $\omega_{b1}\neq\omega_{b3}$. Numerical
solution of Eq.(\ref{eq3.7}) gives dispersion curves for different
thicknesses of the waveguide $a=0.1, 0.3, 0.5, 3~ \mu m$ depicted in
Fig. \ref{fig4}(a), (b), (c) and (d) for both $q>0$ and $q<0$. As we
expect for an asymmetric waveguide the SPP dispersion is
nonreciprocal, namely it depends on the propagation direction, due
to the difference in magnitude of the chiral anomaly in WSMs. Our
results show a tremendous range of frequency
($\Delta\omega\sim\Omega_p$) for unidirectional propagation of SPPs.
This means that in a large range of frequency, which can be tuned
through the chemical potential, SPP is propagated in one direction,
while propagation of the SPP in the backward direction is forbidden.
This property can be broadly used in realizing unidirectional
optical circuits without need to use an external magnetic field. As
we can see from Fig. \ref{fig4}(a), (b), (c) and (d), by decreasing
the thickness of the waveguide a global shift of the bands toward
the lower frequencies is observed. For $q>0$, the lower band starts
from the origin and continuously approaches to the asymptotic
frequency, but the higher band starts from the $\omega=\Omega_p$ and
then tends to its asymptotic value. For $q<0$, the starting points
of the bands are similar to $q>0$, but depending on the value of $a$
the SPP bands may coincide with the bulk plasmon dispersion, which
leads to developing a gap in the SPP dispersion. The gap of the
dispersion in backward direction decreases by decreasing the
thickness of the waveguide and for very small thicknesses gaps are
closed completely. There is no SPP propagation in the frequencies
inside the gap region, but it dose not restrict SPPs unidirectional
propagation regarding the tunability of the gaps by the waveguide
thickness.

\begin{figure}
\centerline{\includegraphics[width=9cm]{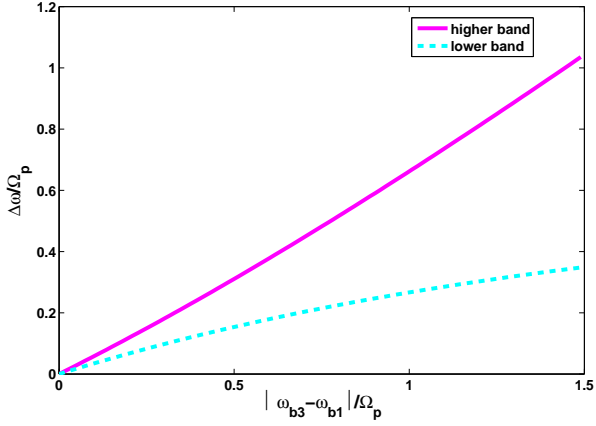}} \caption{
Difference of the asymptotic frequencies of forward and backward
propagation $\Delta\omega=\omega_{+}-\omega_{-}$ as a function of
$|\omega_{b3}-\omega_{b1}|$. Where we have considered
$\omega_{b1}/\Omega_p=0.5$ and the other parameters are similar to
Fig. \ref{fig2}.} \label{fig5}
\end{figure}

To emphasis the tremendous range of frequencies of the
unidirectional SPP propagation we obtain the asymptotic frequencies.
In the nonretarded limit ($|q|\gg k_{0}$), we obtain the asymptotic
frequencies for higher band,
\begin{equation}
\omega_{as}^{v+}=\frac{\sqrt{\varepsilon_{\infty}^{2}\omega_{b1}^{2}+
4\varepsilon_{\infty}
\Omega_{p}^{2}(\varepsilon_{d}+\varepsilon_{\infty})}+
\varepsilon_{\infty}\omega_{b1}}{2(\varepsilon_{d}+\varepsilon_{\infty})}~~~for~q>0
, \label{eq3.11}
\end{equation}
\begin{equation}
\omega_{as}^{v-}=\frac{\sqrt{\varepsilon_{\infty}^{2}\omega_{b3}^{2}+
4\varepsilon_{\infty}
\Omega_{p}^{2}(\varepsilon_{d}+\varepsilon_{\infty})}+
\varepsilon_{\infty}\omega_{b3}}{2(\varepsilon_{d}+\varepsilon_{\infty})}~~~for~
q<0 , \label{eq3.12}
\end{equation}
and for the lower band,
\begin{equation}
\omega_{as}^{v+}=\frac{\sqrt{\varepsilon_{\infty}^{2}\omega_{b3}^{2}+
4\varepsilon_{\infty}
\Omega_{p}^{2}(\varepsilon_{d}+\varepsilon_{\infty})}-
\varepsilon_{\infty}\omega_{b3}}{2(\varepsilon_{d}+\varepsilon_{\infty})}~~~for~q>0
, \label{eq3.13}
\end{equation}
\begin{equation}
\omega_{as}^{v-}=\frac{\sqrt{\varepsilon_{\infty}^{2}\omega_{b1}^{2}+
4\varepsilon_{\infty}
\Omega_{p}^{2}(\varepsilon_{d}+\varepsilon_{\infty})}-
\varepsilon_{\infty}\omega_{b1}}{2(\varepsilon_{d}+\varepsilon_{\infty})}~~~for~q<0
. \label{eq3.14}
\end{equation}
These asymptotic frequencies have shown by thin black lines in Figs.
(\ref{fig4}). Considering the nonretarded limit of the SPPs
dispersion it can be shown that the unidirectional propagation range
grows monotonically by increasing contrast of two WSMs chiral
anomalies. We illustrate aforementioned point by plotting
$\Delta\omega=\omega_{+}-\omega_{-}$ as a function of
$|\omega_{b3}-\omega_{b1}|$ for lower and higher bands. As we can
see in Fig. (\ref{fig5}), for higher band the unidirectional
propagation range grows very faster than the lower band. As a
striking result we observe that a quite huge enhancement of
nonreciprocal effect is accessible by using two different WSM
materials in the slot waveguide. It is worth to mention that the
chemical potentials of WSMs and waveguide thickness can be used as
fine-tuning to acquire a prominent  and robust nonreciprocal effect
in the proposed structure without implementation of an external
magnetic field.
%
%
%
%

%
\begin{figure*}
\centerline{\includegraphics[width=18cm]{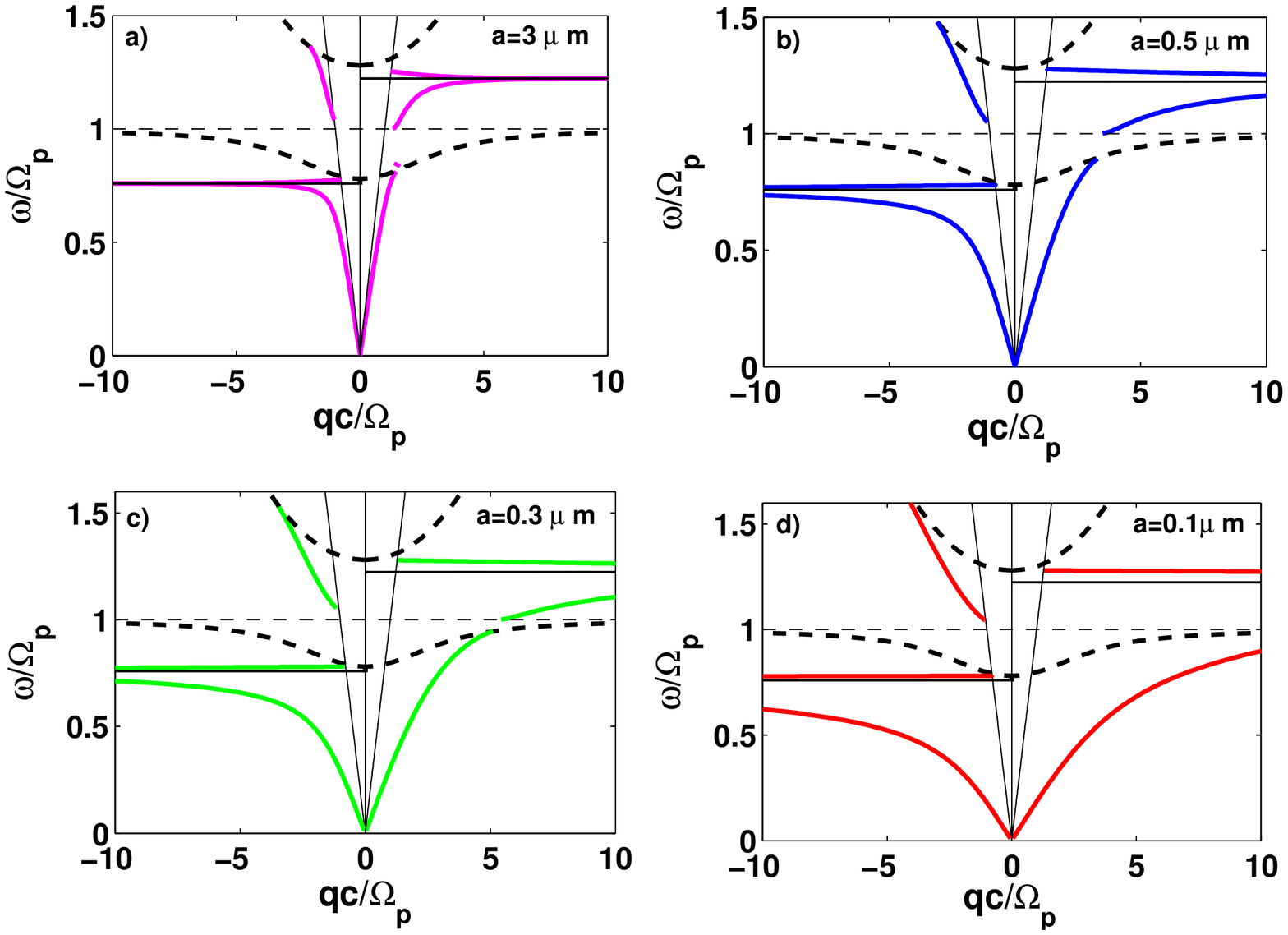}} \caption{SPP
dispersion curve as a function of wave vector for the asymmetric
Voigt-Voigt waveguide with antiparallel $\mathbf{b}$ vectors for
$\omega_{b1}=\omega_{b3}=0.5~ \Omega_p$ and the other parameters
identical with Fig. \ref{fig2}. The black dashed lines show the bulk
plasmon dispersion and the black solid thin lines indicate the
asymptotic frequencies.} \label{fig6}
\end{figure*}
%

Now we turn to the second case, namely difference in the
$\mathbf{b}$ vectors directions. To realize it, we consider vector
of nodes separation in two WSMs points in opposite directions
$\mathbf{b_1}=-\mathbf{b_2}=\mathbf{b}$. For this situation the main
Eq. \ref{eq3.7} reduces to the following one,
\begin{equation}
2{\varepsilon _d}k/\left( {{k_2}{\varepsilon _{w1}}} \right) +
\left[ {1 + {\varepsilon _d}^2{k^2}/\left( {{\varepsilon
_{w1}}^2{k_2}^2} \right)} \right]\tanh ({k_2}a) = 0 ,
 \label{eq3.15}
\end{equation}
Numerical solution of this equation yields the SPP dispersion shown
in Fig. \ref{fig6} as a function of wave vector for different
thicknesses of the waveguide. Our results tend to the SPP dispersion
for a single interface of WSM and dielectric in the Voigt
configuration at the wide waveguide limit. In this limit, for $q>0$
the lower band starts from the origin an end when it intersects by
the bulk plasmon dispersion, but the higher band starts from the
light line above the bulk plasmon frequency and then tends to the
asymptotic value. For $q<0$ the lower band starts from the origin
and continuously approaches to its asymptotic frequency, while the
higher band starts from the light line and continues until it
coincides with the bulk plasmon dispersion. Obviously, the
dispersion is nonreciprocal and there is a range of frequencies with
a unidirectional SPP propagation. As we can see from Fig.
\ref{fig6}, decreasing the thickness of the waveguide leads to
splitting of the higher and lower bands for $q>0$ and $q<0$,
respectively. These splittings are due to the mixing of the SPP
modes localized at two WSM interfaces and these split branches form
a two nearly flat bands very close to the asymptotic frequencies.
These nearly flat bands may be employed in slow light
technology\cite{Hu12}. Furthermore, for $q>0$ decreasing to very
small thicknesses of the waveguide leads to merging of the lower
band with lower split branch of the higher band at
$\omega=\Omega_p$.

In the nonretarded limit, $|q|\gg k_{0}$, SPP dispersion equation
reduces to the following equation,
\begin{equation}
2\varepsilon _d/\left( \varepsilon_{1}-\varepsilon_{b1} \right) +
\left( 1 + \varepsilon_d^2/\left( \varepsilon_{1}-\varepsilon_{b1}
\right)^2 \right)\tanh(|q|a) = 0 .
 \label{eq3.16}
\end{equation}
The asymptotic frequencies are obtained by solving this equation in
the limit of $|q|\rightarrow\infty$, which is identical with the
result for a symmetric Voigt-Voigt configuration given by Eq.
\ref{eq3.10}.
%
\subsection{Surface plasmon polariton in a symmetric Faraday-Faraday waveguide }
%
To study properties of the SPP dispersion in Faraday configuration,
we consider a symmetric waveguide with two identical WSMs in Faraday
configuration. In this case, direction of $\mathbf{b}$ is assumed to
be parallel to the propagation direction $\mathbf{q}$ and both of
them are taken to lie along $y$ axis. Thus, dielectric tensors of
WSMs in the Faraday configuration are given by,
\begin{equation}
\hat{\varepsilon}_{F}(\omega )=\left({\begin{array}{*{20}{c}}
\varepsilon&0& i\varepsilon_{b}\\
0&\varepsilon&0\\-i\varepsilon_{b}&0&\varepsilon
\end{array}} \right) .\begin{array}{*{20}{c}}
\end{array}\label{eq3.17}
\end{equation}
Substituting $\hat{\varepsilon}_{F}(\omega)$ in the wave equation
results in a system of three linear equations resulting to a
$\hat{M}$ matrix for both mediums \textbf{I} and \textbf{III} as,
\begin{equation}
\hat{M}_{F} = \left( {\begin{array}{*{20}{c}}
{q^2 - k^2 - \varepsilon k_{0}^2}&0&{-ik_{0}^2\varepsilon_{b}}\\
0&{-k^2 - \varepsilon k_{0}^2}&{\mp iq{k}}\\
{+ik_{0}^2\varepsilon_{b}}&{\mp iq{k}}&{q^2 - \varepsilon k_{0}^2}
\end{array}} \right) .\label{eq3.18}
\end{equation}
Solutions for decaying wave vectors are obtained by making the
determinant of $\hat{M}$ to be zero. As a result the decaying
constants in both WSMs are given by,
\begin{equation}
k_{\pm}^{2} = k^2 +
k_{0}^2\left(\frac{\varepsilon_{b}^2}{2\varepsilon}\right) \pm
\left[k_{0}^4\frac{\varepsilon_{b}^4}{2\varepsilon^2 }+q^2 k_{0}^2
\frac{\varepsilon_{b}^2}{\varepsilon}\right]^{1/2} ,\label{eq3.19}
\end{equation}
where $k^2 = q^2 - {k_{0}^2}\varepsilon$. So, components of the
electric fields in three regions are expressed by,
\begin{widetext}
\begin{equation}\small{
\left\{\begin{array}{l} \mathbf{E}_1 =
{\begin{array}{*{20}{c}}\left[ {E_{1x}^{+}e^{-k_{+}(z - a/2)} +
E_{1x}^{-}e^{-k_{-}(z - a/2)}},~{\chi_{+}E_{1x}^{+}e^{-k_{+}(z -
a/2)}+ \chi_{-}E_{1x}^{-}e^{-k_{-}(z -
a/2)}},~{\eta_{+}E_{1x}^{+}e^{-k_{+}(z - a/2)} +
\eta_{-}E_{1x}^{-}e^{-k_{-}(z - a/2)}}
\right] ,\end{array}} \\
\mathbf{E}_2 = \left[ {\begin{array}{*{20}{c}} {E_{2x}^{+}e^{+ k_2z}
+ E_{2x}^{-}e^{-k_2z}},&{E_{2y}^{+}e^{+k_2z} +
E_{2y}^{-}e^{-k_2z}},&{\beta(E_{2y}^{+}e^{+k_2z} -
E_{2y}^{-}e^{-k_2z})}\end{array}} \right] ,\\
\mathbf{E}_{3} =  {\begin{array}{*{20}{c}}
\left[{E_{3x}^{+}e^{k_{+}(z + a/2)} + E_{3x}^{-}e^{k_{-}(z +
a/2)}},~{\chi _{+}E_{3x}^{+}e^{k_{+}(z + a/2)} +
\chi_{-}E_{3x}^{-}e^{k_{-}(z + a/2)}},~{\eta_{+}E_{3x}^{+}e^{k_{+}(z
+ a/2)} + \eta_{-}E_{3x}^{-}e^{k_{-}(z + a/2)}} \right]
,\end{array}}
\end{array} \right.}\label{eq3.20}
\end{equation}
\end{widetext}
where $\eta _{\pm} = \frac{E_{iz}}{E_{ix}} = \frac{q^2 - k_{\pm}^2 -
k_{0}^2\varepsilon}{ik_{0}^2\varepsilon_{b}}$, $\chi_{\pm} = \mp
A_{\pm}\left(\frac{qk_{\pm}}{k_{0}^2\varepsilon _{b}}\right)$, with
$A_{\pm}=\frac{q^2 - k_{\pm}^2 - k_{0}^2\varepsilon}{k_{\pm}^2 +
k_{0}^2\varepsilon}$. As the boundary condition, the tangential
components of the electric and magnetic fields must to be matched at
two interfaces. To do this we write the ${E_y}$, ${H_x}$ and ${H_y}$
components of fields in media I and III in terms of ${E_{x}}$ in the
same medium. Application of the boundary conditions results in the
following system of eight equations,
\begin{equation}\small{
\begin{array}{l}
e^{k_2a/2}E_{2x}^{+} + e^{-k_2a/2}E_{2x}^{-} = E_{1x}^{+} + E_{1x}^{-} ,\\
e^{k_2a/2}{E_{2y}^{+}} + {e^{-k_2a/2}}{E_{2y}{-}} =
\left(\frac{-q}{k_{0}^2\varepsilon_{b}}\right)
\left( A_{+}k_{+}E_{1x}^{+} + A_{-}k_{-}E_{1x}^{-}\right) ,\\
e^{k_2a/2}E_{2y}^{+} - e^{-k_2a/2}E_{2y}^{-} =
\left(\frac{qk_2\varepsilon}{k_{0}^2\varepsilon_{b}
\varepsilon_d}\right)\left(A_{+}E_{1x}^{+}
+ A_{-}E_{1x}^{-}\right) ,\\
e^{k_2a/2}E_{2x}^{+} - e^{-k_2a/2}E_{2x}^{-} = -
\frac{k_{+}}{k_2}E_{1x}^{+} - \frac{k_{-}}{k_2}E_{1x}^{-} ,\\
e^{-k_2a/2}E_{2x}^{+} + e^{k_2a/2}E_{2x}^{-} = E_{3x}^{+} + E_{3x}^{-} ,\\
e^{-k_2a/2}E_{2y}^{+} + e^{k_2a/2}E_{2y}^{-} =
\left(\frac{q}{k_{0}^2\varepsilon_{b}}\right)
\left(A_{+}k_{+}E_{3x}^{+} + A_{-}k_{-}E_{3x}^{-}\right) ,\\
e^{-k_2a/2}E_{2y}^{+} - e^{k_2a/2}E_{2y}^{-} =
\left(\frac{qk_2\varepsilon}{k_{0}
^2\varepsilon_{b}\varepsilon_d}\right)
\left(A_{+}E_{3x}^{+} + A_{-}E_{3x}^{-}\right) ,\\
e^{-k_2a/2}E_{2x}^{+} - e^{k_2a/2}E_{2x}^{-} =
\frac{k_{+}}{k_2}E_{3x}^{+} + \frac{k_{-}}{k_2}E_{3x}^{-} .
\end{array}}\label{eq3.21}
\end{equation}
Setting the determinant of its coefficient matrix to zero yields the
dispersion relation as,
\begin{widetext}
\begin{equation}
\small{\begin{array}{l} \left[\left(2\varepsilon_d\varepsilon
{k_2}^2\left(2k_- k_+({A_-}^2 + {A_+}^2) - {A_-}{A_+}({k_-} +
{k_+})^2\right) + (-2{A_-}{A_+} \left({k_2}^2 +
{k_-}{k_+}\right)\left(\varepsilon^2{k_2}^2 +
\varepsilon_d^2{k_-}{k_+}\right) \right.\right.\\\left.\left.
+{A_-}^2\left(\varepsilon^2{k_2}^2 +
\varepsilon_d^2{k_-}^2\right)({k_2}^2 + {k_+}^2) +
{A_+}^2\left(\varepsilon^2{k_2}^2 + \varepsilon
_d^2{k_+}^2\right)({k_2}^2 + {k_-}^2)){\tanh^2(k_2
a)}\right){\cosh^2(k_2a)} \right.
\\\left. +{k_2}(2{A_-}{A_+}{\varepsilon_d}\varepsilon{k_2}{({k_-} -
{k_+})^2} + ({A_-} - {A_+})\left({\varepsilon_d}({A_-}{k_-} -
{A_+}{k_+}) - \varepsilon ({A_+}{k_-} +
{A_-}{k_+})\right)\left(\varepsilon{k_2}^2 +
{\varepsilon_d}{k_-}{k_+}\right)\cosh(2{k_2}a))\right]=0~ .
\end{array}}\label{eq3.22}
\end{equation}
\end{widetext}
It can be shown that in the nonretarded limit, Eq.(\ref{eq3.22})
reduces to,
\begin{equation}
{2{\varepsilon _d}\varepsilon  + \left( {{\varepsilon _d}^2 +
{\varepsilon ^2}} \right){\mathop{\rm tanh}\nolimits} \left( {qa}
\right)}=0 ,\label{eq3.23}
\end{equation}
and in this case, the asymptotic frequency reads,
$\omega_{as}^{F}=\Omega_{p} \frac{{\sqrt {{\varepsilon _\infty }}
}}{{\sqrt {{\varepsilon _d} + {\varepsilon _\infty }} }}$.

\begin{figure*}
\centerline{\includegraphics[width=18cm]{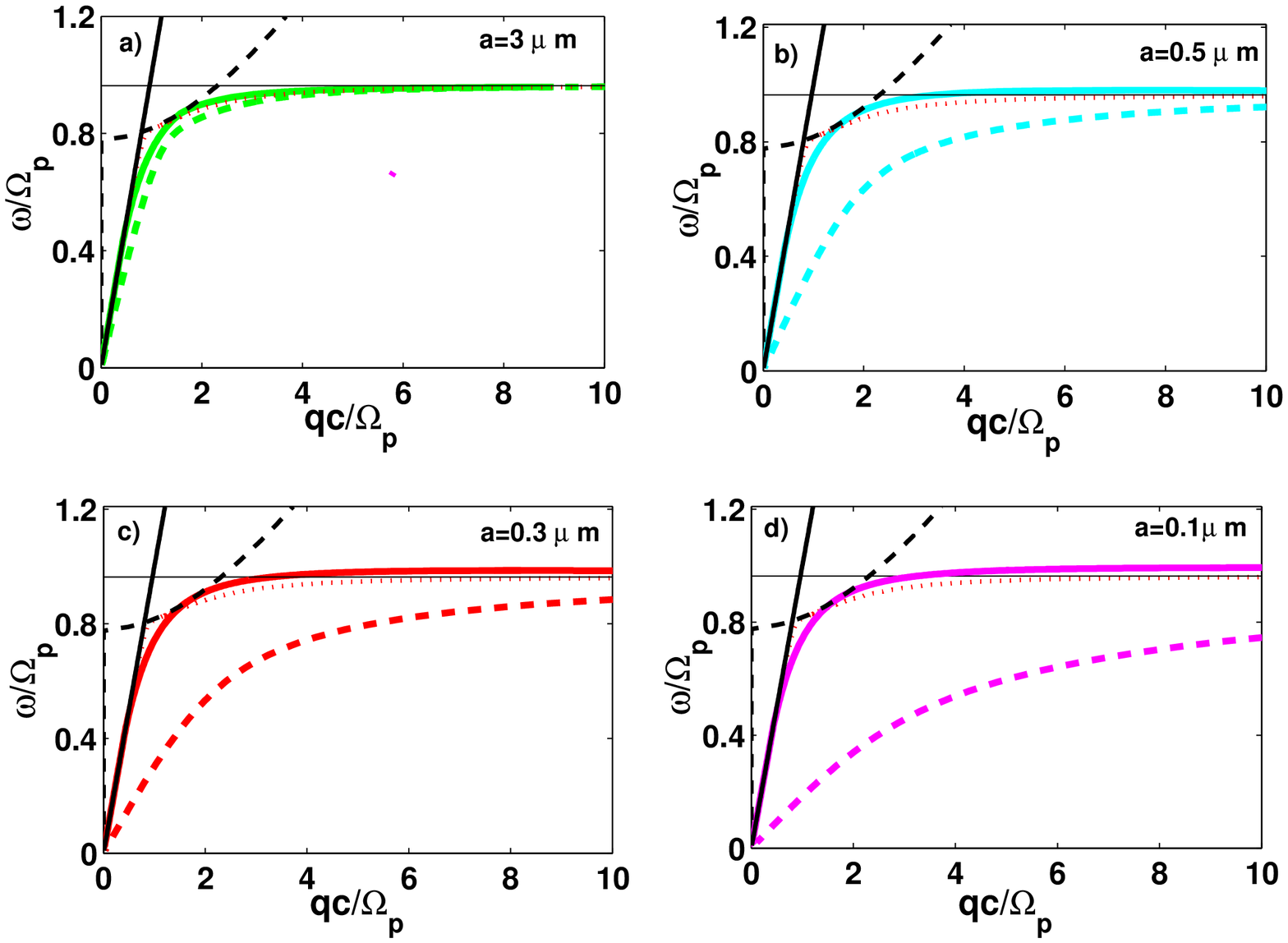}} \caption{The
dispersion curves of the symmetric slot waveguide in the Faraday
configuration with different thickness $a=3, 0.5, 0.3$ and $0.1~ \mu
m $. The other parameters are identical with Fig. \ref{fig2}.}
\label{fig7}
\end{figure*}

\begin{figure*}
\centerline{\includegraphics[width=18cm]{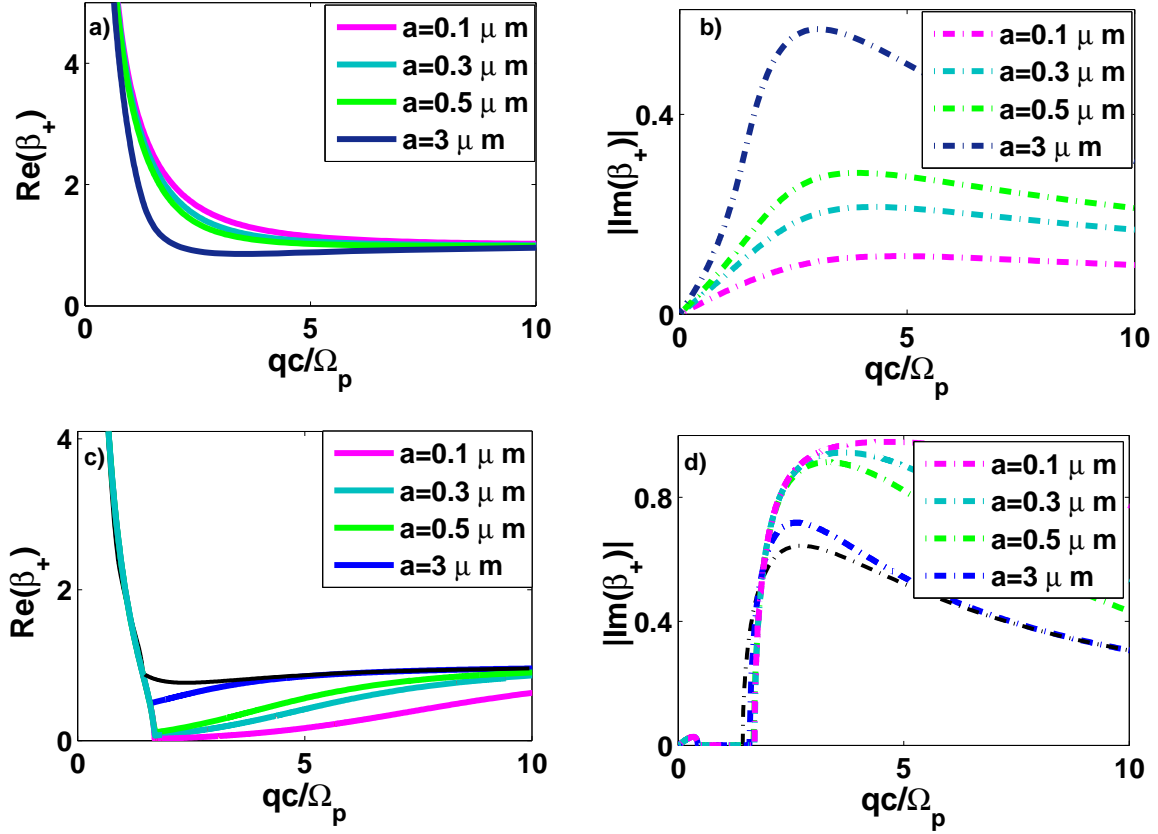}} \caption{a) Real
and b) imaginary parts of the reduced decay constants Of lower band,
c) real and b) imaginary parts of the reduced decay constants of
higher band versus SPP wave vector in the symmetric slot waveguide
with Faraday configuration for different thickness $a=3, 0.5, 0.3$
and $0.1~ \mu m $.} \label{fig8}
\end{figure*}

Dispersion curves of the symmetric Faraday- Faraday waveguide has
been plotted in Fig.\ref{fig7} for different thicknesses of the
dielectric $a=3, 0.5, 0.3, 0.1~ \mu m $ as a function of SPP wave
vector. As it can be seen from Fig.\ref{fig7}, the SPP dispersion
composes of two bands, one is very close to the SPP dispersion for a
single interface with Faraday configuration, denoted by a red dotted
curve, and the other one lies below it. For a wide waveguide these
two bands merge into a one band coinciding with the result for a
single interface. Decreasing the thickness of the waveguide leads to
shifting of the lower band to the lower frequencies due to the
mixing of the SPP modes of two interfaces. Further inspection
reveals that the SSP modes are generalized surface waves, i.e. the
decaying constants $k_{+}$ and $k_{-}$ are complex conjugates of
each other\cite{Wallis74}. The Real and imaginary parts of the
reduced decay constants $\beta_+ = k_+/q$ has been plotted as a
function of the SPP wave vector for lower and higher bands of SPP
dispersion in Figs. \ref{fig8} (a), (b) and Figs. \ref{fig8} (c),
(d), respectively. As it can be seen, the real part of $\beta_+$
decreases with increasing of the wave vector and at large wave
vectors it approaches to an asymptotic value for both bands. The
imaginary part of the $\beta_+$ for both bands has a maximum in the
intermediate wave vectors before approaching the asymptotic value at
large wave vectors. The higher band has an interesting property that
it has been composed of normal SPP modes (with vanishing imaginary
part of the decaying constant) and generalized SPP modes.
%
%
%
%
%
\begin{figure*}
\centerline{\includegraphics[width=18cm]{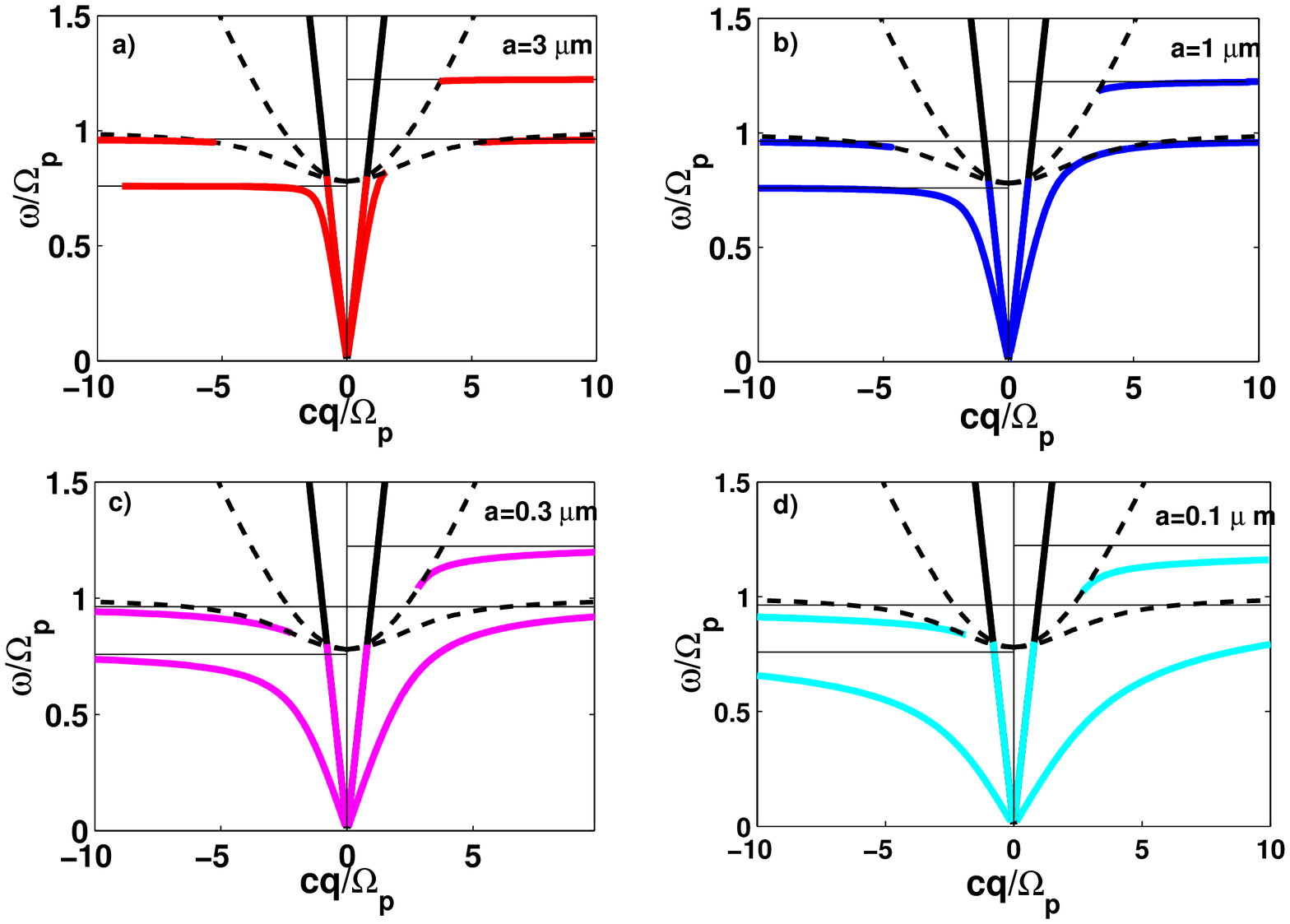}} \caption{Surface
plasmon polariton of asymmetry Voigt-Faraday waveguide for different
thickness of dielectric layer $a=3, 1, 0.3$ and $0.1~ \mu m $. The
other parameters are identical with Fig. \ref{fig2}. } \label{fig9}
\end{figure*}
%
\subsection{Surface plasmon polariton in an asymmetric Voigt-Faraday waveguide}
%
As an exotic structure, we consider an asymmetric waveguide
comprising of two semi-infinite WSMs in two different Voigt and
Faraday configurations. Since vector $\mathbf{b}$, which plays a
role similar to an external magnetic field in conventional
semiconductor magneto optic materials, has different directions in
two WSMs due to their different configurations, so the mentioned
structure resembles to a waveguide placed in an external magnetic
field having different directions in mediums I and III. Since chiral
anomaly is an intrinsic property of the WSMs, the considered
structure is experimentally achievable, while realizing such
structure using external magnetic field may be a challenging task.
This configuration exploits features of both Voigt and Faraday
configuration simultaneously. It has been assumed that mediums I and
III are in Voigt and Faraday configuration with dielectric tensors
given by Eq. (\ref{eq3.1}) and Eq. (\ref{eq3.17}), respectively. The
decay constant for the Voigt configuration (medium I) is given by
Eq. (\ref{eq3.3}) and for Faraday configuration (medium III) by Eq.
(\ref{eq3.19}). Therefore, components of the electric field in
different mediums are expressed as,
\begin{widetext}
\begin{equation}\small{ \left\{ \begin{array}{l} {E_1}(z) =
\left[ {\begin{array}{*{20}{c}} {{E_{1x}}{e^{ - {k_{1}^{+}}}}(z -
a/2)},&{{E_{1y}}{e^{ - {k_{1}^{-}}}(z - a/2)}},&{\beta_{1}
{E_{1y}}{e^{ - {k_{1}^{-}}}(z - a/2)}}\end{array}} \right] ,\\
{E_2}(z) = \left[ {\begin{array}{*{20}{c}} {{E_{2x}^{+}}{e^{ +
{k_2}z}} + {E_{2x}^{-}}{e^{ - {k_2}z}}},&{{E_{2y}^{+}}{e^{ +
{k_2}z}} + {E_{2y}^{-}}{e^{ - {k_2}z}}},&{\beta_{2}
({E_{2y}^{+}}{e^{ + {k_2}z}} - {E_{2y}^{-}}{e^{ - {k_2}z}})}
\end{array}} \right] ,\\ {E_3}(z) = {\begin{array}{*{20}{c}}\left[
{{E_{3x}^{+}}{e^{{k_{3}^{+}}(z + a/2)}} +
{E_{3x}^{-}}{e^{{k_{3}^{-}}}(z + a/2)}},~{{\chi
_+}{E_{3x}^{+}}{e^{{k_{3}^{+}}(z + a/2)}} + {\chi
_-}{E_{3x}^{-}}{e^{{k_{3}^{-}}(z +
a/2)}}},~{{\eta_+}{E_{3x}^{+}}{e^{{k_{3}^{+}}(z + a/2)}} +
{\eta_-}{E_{3x}^{-}}{e^{{k_{3}^{-}}(z + a/2)}}} \right] ,
\end{array}} \end{array} \right.}\label{eq3.24}
\end{equation}
\end{widetext}
where $\beta_{1} = \frac{E_{1z}}{E_{1y}} =  - \frac{(-iq{k_{1}^{-}}
+{k_{0}^2}\varepsilon_{b})}{q^2 - {k_{0}^2}\varepsilon}$ ,
$\beta_{2} = \frac{E_{2z}}{E_{2y}} =  - \frac{iq{k_2}}{q^2 -
{k_{0}^2}{\varepsilon _d}}$ , ${\eta_{\pm}} = \frac{E_{3z}}{E_{3x}}
 = \frac{q^2 - {k_{3}^{\pm}}^{2} - {k_{0} ^2}\varepsilon }{{ik_{0}
^2}\varepsilon _{b}}$ and ${\chi_{\pm}} = \frac{E_{3y}}{E_{3x}} =
{A_{\pm }}(\frac{ + iq{k_{3}^{ \pm }}}{{{ik_{0} ^2}\varepsilon
_{b}}})$ with ${A_{\pm}} = (\frac{q^2 - {k_{3}^{\pm}}^2 - {k_{0}
^2}\varepsilon}{ {k_{3}^{\pm }}^2 + {k_{0} ^2}\varepsilon })$. Again
by employing boundary conditions at two interfaces $z=\pm a/2$,
eight linear equations are obtained for eight unknown amplitudes of
electric filed ($E_{1x}, E_{1y}, E_{2x}^{+}, E_{2y}^{+}, E_{2x}^{-},
E_{2y}^{-}, E_{3y}^{+}, E_{3y}^{-}$). Using the same procedure
explained in the previous sections the following dispersion relation
can be obtained for this structure,
\begin{widetext}
\begin{equation}\begin{array}{l}
\left[-A_{+}\left(\varepsilon_d k_2\left(\varepsilon
k_{1-}+\varepsilon_A k_{3+}\right) + \varepsilon \varepsilon _A
k_{d}^{2}T + \varepsilon_d^{2} k_{1-}k_{3+}T \right)\left(
k_2\left(k_{1+} + k_{3-} \right) + k_2^{2}T + k_{1+}k_{3-}T \right)
\right.\\\left. + A_{-}\left( \varepsilon_d k_2\left(\varepsilon
k_{1-} + \varepsilon_A k_{3-} \right) + \varepsilon \varepsilon_A
k_2^{2}T + \varepsilon_d^{2} k_{1-}k_{3-}T\right)
\left(k_2\left(k_{1+} + k_{3+} \right) + k_2^{2}T + k_{1+}k_{3+}T
\right)\right]\cosh^2 \left(k_{2}a \right)=0 , \label{eq3.25}
\end{array}
\end{equation}
\end{widetext}
where $T=tanh(k_{2} a)$. In the nonretarded limit, Eq(\ref{eq3.25})
reduces to,
\begin{equation}
\varepsilon_d\left( -2i\varepsilon + \varepsilon_{b}\right) + \left[
\left( - i\varepsilon + \varepsilon_{b} \right)\varepsilon -
i\varepsilon_d^{2} \right]\tanh(q a) =0 .\label{eq3.26}
\end{equation}
For $aq\gg 1$, we end up with the following asymptotic frequencies,
\begin{equation}
\begin{array}{l}
\omega_{as}^{f} = \Omega_{p} \frac{\sqrt
{\varepsilon_{\infty}}}{\sqrt{\varepsilon_d + \varepsilon_{\infty}}}\\
\omega_{as}^{v}  = \frac{{{\Omega_{p} \omega _{b}} \pm \sqrt
{4{\varepsilon _\infty
}\Omega_{p}^{2}(\varepsilon_{\infty}+\varepsilon _d)+
\Omega_{p}^{2}\omega_{b}^2}}}{{2\left( {{\varepsilon
_\infty}+{\varepsilon _d }} \right)}}
\end{array}\label{eq3.27}
\end{equation}
Fig. (\ref{fig9}) shows the SPP dispersion for the asymmetric
Voigt-Faraday waveguide for the dielectric thicknesses $a=3, 1, 0.3,
0.1~ \mu m$. It is remarkable that the nonreciprocal effect - i.e.,
nonequivalent dispersion for positive and negative wave vectors - is
observed in this configuration. As it is evident from Fig.
(\ref{fig9}), for a wide waveguide ($a=3 \mu m$) the dispersion
curves are nearly identical with the results for a single interface
between WSM and dielectric with Voigt and Faraday configurations
\cite{Hofmann16}. The bands with Faraday character have two branches
below the bulk plasmon frequency and are reciprocal, in contrast the
bands possessing Voigt character are nonreciprocal with two higher
and lower brands for forward propagation and a continues band for
backward direction. In general the SPP modes above the plasmon
frequency propagate unidirectional, while the SPP modes below the
$\Omega_p$ are nonreciprocal but bidirectional. Decreasing the
waveguide thickness leads to shifting the bands toward lower
frequencies. An interesting result is merging the lower Voigt band
with a branch of Faraday band close to $\Omega_p$ for $q>0$ with
decreasing the waveguide thickness, but the higher Voigt band
retains its gap where it coincides with bulk plasmon dispersion. It
leads to a continues band which posses the Voigt and Faraday
characteristic simultaneously.

%
\section{conclusion}\label{S3}
%
In conclusion we have studied the SPP dispersion in a slot waveguide
constructed by two WSMs connected via a dielectric layer. Here, we
have considered novel and exotic configurations for SPP propagation
due to the intrinsic topological properties of the WSMs. The
symmetric Voigt-Voigt waveguide shows a reciprocal SPP dispersion.
But, we showed that we can retrieve the nonreciprocal and
unidirectional SPP propagation in Voigt configuration by breaking
the symmetry of the structure via generating a contrast in chiral
anomaly magnitude or its direction in two WSMs. It is remarkable
that we observe a tremendous range of frequency for unidirectional
SPP propagation. Furthermore, we showed that this unidirectional
propagation can be fine-tuned by the waveguide thickness and the
chemical potentials of two WSMs. Moreover, to complete our study we
investigated the Faraday-Faraday waveguide which shows the
reciprocal SPP dispersion with two bands below the bulk plasmon
frequency. As an hybrid structure we studied the SPP dispersion in
the Voigt-Faraday waveguide. We interestingly found that it shows a
unidirectional SPP propagation above the bulk plasmon frequency,
while it shows a nonreciprocal but bidirectional SPP dispersion
below the bulk plasmon frequency. In summery, we observed a
tremendous unidirectional SPP propagation in the structures
introduced in this study. We believe that our results can be
observed experimentally and they may be useful in creating
unidirectional optical devices and in the slow light technology.


%

\end{document}